\documentclass[11pt,nofootinbib,floatfix]{revtex4}
\usepackage{mathrsfs}
\usepackage{graphicx}
\usepackage{amsmath}
\usepackage{amsfonts}
\usepackage{amssymb}
\usepackage{color}

\usepackage{epsfig}
\usepackage{CJK}
\usepackage{graphicx}
\usepackage{epsfig}
\usepackage{eepic}
\usepackage{bbm}
\usepackage{dcolumn}
\usepackage{bm}
\usepackage{ulem}
\usepackage{slashbox}
\usepackage{multirow}
\usepackage{slashed}

\newcommand{\omits}[1]{}

\def\bc{\begin{center}}
\def\nno{\nonumber}
\def\ec{\end{center}}
\def\be{\begin{eqnarray}}
\def\ee{\end{eqnarray}}

\definecolor{dyellow}{rgb}{1.,0.8,.0}
\definecolor{myblue}{rgb}{.1,.1,.7}
\definecolor{dcyan}{rgb}{.0,.6,.6}
\definecolor{cyan}{rgb}{0.4,1.0,1.0}
\definecolor{dmagenta}{rgb}{0.6,0.0,0.6}
\definecolor{brown}{rgb}{0.6,0.2,0.}
\definecolor{darkblue}{rgb}{.0,.0,0.5}
\definecolor{darkred}{rgb}{0.75,0.0,0.0}
\definecolor{orange}{rgb}{1.,.6,.0}
\definecolor{dorange}{rgb}{0.8,.4,.0}
\definecolor{green}{rgb}{0.0,1.0,0.0}
\definecolor{darkgreen}{rgb}{0.0,0.6,0.0}
\definecolor{purple}{rgb}{.4,.0,.4}
\definecolor{lightgrey}{rgb}{0.7, 0.7, 0.7}
\definecolor{grey}{rgb}{0.4, 0.4, 0.4}

\def\red{\color{red}}

\begin{document}


\title{CFT dual of charged AdS black hole in the large dimension limit}

\author{Er-Dong Guo$^{1,3}$} \email{guoerdong@itp.ac.cn}
\author{Miao Li$^{2}$} \email{limiao9@mail.sysu.edu.cn}
\author{Jia-Rui Sun$^{2}$} \email{sunjiarui@mail.sysu.edu.cn}

\affiliation{${}^1$State Key Laboratory of Theoretical Physics, Institute of Theoretical Physics, Chinese Academy of Science, Beijing 100190, China}
\affiliation{${}^2$School of Physics and Astronomy, Sun Yat-Sen University, Guangzhou 510275, China}
\affiliation{${}^3$Kavli Institute for Theoretical Physics China, Chinese Academy of Sciences, Beijing, 100190}



\begin{abstract}
We study the dual CFT description of the $d+1$-dimensional Reissner-Nordstr\"om-Anti de Sitter (RN-AdS$_{d+1}$) black hole in the large dimension (large $d$) limit, both for the extremal and nonextremal cases. The central charge of the dual CFT$_2$ (or chiral CFT$_1$) is calculated for the near horizon near extremal geometry which possess an  AdS$_2$ structure. Besides, the $Q$-picture hidden conformal symmetry in the nonextremal background can be naturally obtained by a probe charged scalar field in the large $d$ limit, without the need to input the usual limits to probe the hidden conformal symmetry. Furthermore, an new dual CFT description of the nonextremal RN-AdS$_{d+1}$ black hole is found in the large $d$ limit and the duality is analyzed by comparing the entropies, the absorption cross sections and the retarded Green's functions obtained both from the gravity and the dual CFT sides.
\end{abstract}


\maketitle
\tableofcontents

\section{Introduction}
One of the significant achievements of string theory is that the Bekenstein-Hawking entropy~\cite{Bekenstein:1973ur,Hawking:1976de} of some kinds of black holes could be reproduced by counting microstates of the dual CFTs~\cite{Strominger:Black hole entropy from near horizon microstates,Strominger:Microscopic origin of the Bekenstein-Hawking entropy}. The precise match gives strong support that string theory is a correct quantum theory of gravity.

Later it was realized that, any consistent, unitary quantum theory containing the special kinds of black holes which have the near-horizon AdS structure must reproduce the black hole thermal entropy. This fact is understood easily because the universal area-entropy law should not depend on the exact UV behavior of the quantum gravity.

In recent period of time, the so-called Kerr/CFT correspondence~\cite{Guica:2008mu,Lu:2008jk,Bredberg:2009pv,Castro:2009jf,Hartman:2008pb,Cvetic:2009jn,
Chow:2008dp,Lu:2009gj,Castro:2008ms,Compere:2010uk} and the Reissner-Nordstr\"om (RN)/CFT duality~\cite{Garousi:2009zx,Chen:2010bs,Chen:2009ht,Chen:2012pt,Chen:2012ps} has been widely and deeply investigated and there have been many generalizations and extensions of the Kerr/CFT correspondence. \omits{Using this tool(AdS/CFT), we can also study many strong coupling system which can not be calculated using the traditional method~\cite{Lin:2015acg}.}For the Kerr and the RN black holes, a certain AdS structure appears when taking the near horizon near extremal limit. By analyzing the asymptotic symmetry of the background spacetime and calculating the boundary stress tensor of the two dimensional effective action, the central charge of the dual CFT can be obtained~\cite{Guica:2008mu,Castro:2008ms,Chen:2009ht}. For some nonextremal black holes, the conformal symmetries are not manifest because of lacking AdS structure, but they can be probed by a scalar field in a way that the equation of motion of the scalar field reveals the conformal symmetry inherited from the black hole background geometry, both in the angular momentum $J$-picture for rotating black holes~\cite{Castro:2010fd,Chen:2010xu,Becker:2010dm,Chen:2010bh} and in the charge $Q$-picture for (non)rotating charged black holes~\cite{Chen:2010as,Chen:2010yu,Chen:2010ywa}. See the recent review papers on Kerr/CFT related studies~\cite{Bredberg:2011hp,Compere:2012jk} for more references therein.

According to recent studies of Emparan, in the large $d$ limit, a wide class of non-extremal black holes has a universal near horizon limit, the limiting manifold is the product of a two dimensional and a $d-2$ dimensional compact manifold, which resembles the decoupling limit in the large $N$ limit~\cite{Emparan:2013xia,Emparan:2013moa,Emparan:2014aba}. Thus if we consider the large $d$ limit of a RN-AdS$_{d+1}$ black hole, it is expected that the near horizon near extremal limit of the two-dimensional sub-manifold may contain some new AdS$_2$ structure (which represents a different limit of the original theory), and then its dual CFT description can be studied using the conventional approach.

In this paper, we investigate the holographic CFT dual to the RN-AdS$_{d+1}$ black hole in the large $d$ limit, both in the near extremal and the nonextremal cases. For the extremal case, we calculate the right hand central charge of the dual CFT by analyzing the asymptotic symmetries. While for the nonextremal case, we use similar method applied in probing the charge $Q$-picture hidden conformal symmetry and determine the temperatures and conformal dimensions of the left- and right-hand sectors of the dual CFT. We show that, for the EoM of the probe charged scalar field, when $d$ goes to infinity, the low frequency limit and the low momentum limit can be automatically achieved without the need to impose further conditions required in the conventional method, and the radial EoM is just the eigen equation of the Casimir operator of the $SL(2,R)$ algebra. Interestingly, the results indicate that the dual CFT obtained by probing the hidden conformal symmetry is different from that gained in the near extremal case in the large $d$ limit. We also compute the absorption cross section and retarded Green's functions of the probe charged scalar field and show that they agree with the result of the dual CFT. Therefore, the large $d$ limit not only reveals new dual CFT pictures but also offers an useful tool to probe the hidden conformal symmetry of certain black holes, which will provide us new and nontrivial information about the holographic description of black holes in such limit.

This paper is arranged as follows. We review some basic properties of the geometric background and then derive the near-extremal near-horizon geometry of the large $d$ AdS$_{d+1}$ black hole in Sec.\ref{sec2}. In Sec.\ref{secASG}, we analyze the asymptotical symmetries of the near horizon near extremal background in the large $d$ limit and obtain the central charge of the dual CFT. In Sec.\ref{sechidden}, the probe charged scalar field is added into the nonextremal background, then the hidden conformal symmetry and the scattering process is discussed. The conclusion is drawn in Sec.\ref{seccon}.

\section{The RN-AdS$_{d+1}$ black hole}\label{sec2}
The $d+1$ dimensional Einstein-Maxwell theory has the action as
\begin{equation}\label{actionEM}
I = \frac{1}{16 \pi G_{d+1}} \int d^{d+1}x \sqrt{-g} \left( R + \frac{d(d-1)}{L^2} - \frac{L^2}{g_s^2} F_{\mu\nu} F^{\mu\nu} \right),
\end{equation}
its dynamical equations are
\begin{eqnarray}
R_{\mu\nu} - \frac12 g_{\mu\nu} R - \frac{d(d-1)}{2L^2} g_{\mu\nu} &=& \frac{L^2}{2g_s^2} \left( 4 F_{\mu\lambda} F_{\nu}{}^\lambda - g_{\mu\nu} F_{\alpha\beta} F^{\alpha\beta} \right),
\nonumber\\
\partial_\mu \left( \sqrt{-g} F^{\mu\nu} \right) &=& 0,
\end{eqnarray}
which admit the Reissner-Nordstr{\"o}m-Anti de Sitter (RN-AdS$_{d+1}$) black hole solution
\begin{eqnarray} \label{RNAdS}
ds^2 &=& -f(r) dt^2 + \frac{dr^2}{f(r)} + r^2 d\Omega_{d-1}^2,
\nonumber\\
A &=& \mu \left( 1 - \frac{r_+^{d-2}}{r^{d-2}} \right) dt,
\end{eqnarray}
with
\begin{equation}
f(r) = 1 - \frac{M}{r^{d-2}} + \frac{Q^2}{r^{2d-4}}+\frac{r^2}{L^2}, \qquad {\rm and} \qquad \mu = \sqrt{\frac{d-1}{2(d-2)}} \frac{g_s Q}{L r_+^{d-2}},
\end{equation}
where $r_+$ is the radius of the outer horizon, $\mu$ is the chemical potential with dimension ${\rm dim}[\mu] = {\rm length}^{-1}$, $M$ and $Q$ are related to the ADM mass and physical charge of the RN-AdS$_{d+1}$ black hole as
\be
M_{\rm ADM}=\frac{(d-1)M\Omega_{d-1}}{16\pi G_{d+1}}\quad {\rm and}\quad Q_e=\frac{\sqrt{2(d-1)(d-2)}Q\Omega_{d-1}}{8\pi G_{d+1}g_s}.
\ee
The condition $f(r_+) = 0$ gives
\be M=r_+^{d-2}+\frac{Q^2}{r_+^{d-2}}+\frac{r_+^d}{L^2}\ee
(which is the Smarr like relation related to the first law of thermodynamics of the black hole) and the temperature $T$ and entropy volume density $s$ of the black hole are respectively
\begin{equation}
T = \frac{r_+ d}{4\pi L^2} \left( 1 -\frac{(d-2)}{d}\frac{L^2 (Q^2-r_+^{2d-4}) }{r_+^{2d-2}} \right) \qquad {\rm and} \qquad s = \frac{r_+^{d-1}}{4G_{d+1}}.
\end{equation}
Besides, the first law of thermodynamics of the dual boundary quantum field is
\begin{equation}
\delta \epsilon = T \delta s + \mu \delta\rho,
\end{equation}
where the energy density and the charge density are respectively
\be \epsilon = \frac{M_{\rm ADM}}{\Omega_{d-1}} =\frac{(d-1) M}{16 \pi G_{d+1} }\quad {\rm and}\quad \rho = \frac{Q_e }{\Omega_{d-1}} =\frac{\sqrt{2(d-1)(d-2)} Q}{8 \pi G_{d+1} g_s }.
\ee

\subsection{RNAdS black hole in large $d$ limit}
Define $\varrho\equiv \frac{r^{d-2}}{M}$, then
\be dr^2=\frac{r^2 d\varrho^2}{(d-2)^2\varrho^2},\quad {\rm and}\quad
f(r)=f(\varrho)=1-\frac{1}{\varrho}+\frac{Q^2}{M^2}\frac{1}{\varrho^2}+\frac{(\varrho M)^{\frac{2}{d-2}}}{L^2}.\ee
And eq.(\ref{RNAdS}) becomes
\be \label{RNAdS2} ds^2&=&-\left(1-\frac{1}{\varrho}+\frac{Q^2}{M^2}\frac{1}{\varrho^2}+\frac{(\varrho M)^{\frac{2}{d-2}}}{L^2}\right)dt^2+\frac{\varrho^{\frac{2}{d-2}} M^{\frac{2}{d-2}}d\varrho^2}{(d-2)^2\varrho^2\left(1-\frac{1}{\varrho}+\frac{Q^2}{M^2}\frac{1}{\varrho^2}+\frac{(\varrho M)^{\frac{2}{d-2}}}{L^2}\right)}+\varrho^{\frac{2}{d-2}} M^{\frac{2}{d-2}}d\Omega_{d-1}^2,\nno\\
A &=& \mu \left( 1 - \frac{\varrho_+}{\varrho} \right) dt,
\ee
and $\sqrt{-g}=\frac{ M^{\frac{d}{d-2}}}{(d-2)}\varrho^{\frac{2}{d-2}} \sqrt{{\rm det}\bar{g}_{\theta_i\theta_i}}$, where $g_{\theta_i\theta_i}=\bar{g}_{\theta_i\theta_i}/r^2$. Note that for finite $r>r_+$, taking $d\rightarrow \infty$, then
\be \varrho^{\frac{2}{d-2}}\rightarrow \varrho^0=1,\ee
which means that the finite $r$ region is always driven to the fixed point $r^{d-2}\rightarrow M\equiv r_o^{d-2}$. Meanwhile, the metric (\ref{RNAdS2}) reduces to
\be \label{RNAdS3} ds^2&=&-\left(1-\frac{1}{\varrho}+\frac{Q^2}{M^2}\frac{1}{\varrho^2}+\frac{r_o^2}{L^2}\right)dt^2+\frac{r_o^2
d\varrho^2}{(d-2)^2\varrho^2\left(1-\frac{1}{\varrho}+\frac{Q^2}{M^2}\frac{1}{\varrho^2}+\frac{r_o^2}{L^2}\right)}
+r_o^2 d\Omega_{d-1}^2,\nno\\
A &=& \mu \left( 1 - \frac{\varrho_+}{\varrho} \right) dt.
\ee
The black hole has new inner and outer radius as
\be
\varrho_\pm =\frac{1\pm \sqrt{1-\frac{4k Q^2}{M^2}}}{2k} \quad {\rm with}\quad k\equiv 1+\frac{r_o^2}{L^2}.
\ee
In the extremal case we have $M^2=4 k Q^2=r_o^{2d-4}$, the background geometry also contains the AdS$_2$ structure in the near horizon (the horizon radius are different with the black holes at finite $d$) near extremal limit. Eq.(\ref{RNAdS3}) can also be written as
\be \label{RNAdS4} ds^2&=&\frac{r_o^2}{(d-2)^2}\left(-\left(1-\frac{1}{\varrho}+\frac{Q^2}{M^2}\frac{1}{\varrho^2}+\frac{r_o^2}{L^2}\right)d\tilde{t}^2
+\frac{d\varrho^2}{\varrho^2\left(1-\frac{1}{\varrho}+\frac{Q^2}{M^2}\frac{1}{\varrho^2}+\frac{r_o^2}{L^2}\right)}
+(d-2)^2 d\Omega_{d-1}^2\right),\nno\\
A &=& \tilde{\mu} \left( 1 - \frac{\varrho_+}{\varrho} \right) d\tilde{t},
\ee
where $t=\frac{r_o \tilde{t}}{(d-2)}$, $\tilde{\mu}=\bar{\mu}r_o$ and $\bar{\mu}=\frac{\mu}{(d-2)}$.

Considering the condition
\be
k-\frac{1}{\varrho_{+}}+\frac{Q^{2}}{M^{2}} \frac{1}{\varrho_{+}^{2}}=0
\ee
we have
\be
M=\frac{Q}{\sqrt{\varrho_+-\varrho_+^2 k}},\quad{\rm or}\quad r_o=\left(\frac{Q}{\sqrt{\varrho_+-\varrho_+^2 k}}\right)^{\frac{1}{d-2}}.
\ee
In addition, the temperature associated with the dimensionless time coordinate $\tilde{t}$ and the entropy of the RNAdS$_{d+1}$ black hole in the large $d$ limit are respectively
\be
T=\frac{k(\varrho_+-\varrho_-)}{4\pi \varrho_+}\quad {\rm and}\quad S=\frac{r_o^{d-1}\Omega_{d-1}}{4G_{d+1}}=\frac{1}{4G_2}.
\ee

\subsection{Near horizon near extremal geometry}
The near horizon geometry of the large $d$ limit RN-AdS black hole contains an AdS$_2$ structure naturally.
The procedure of taking limit is as follows,
\be\label{scaling}
\varrho \to \varrho_{+}+\epsilon \tilde{\varrho},\qquad
\tilde{t} \to \frac{\tilde{\tau}}{\epsilon}, \qquad
\varrho_+ - \varrho_- \to \frac{\epsilon}{k},
\ee
then the near extremal near horizon geometry is obtained as
\be\label{nhneld}
ds^2&=&\frac{r_o^2}{k(d-2)^2}\left(-\frac{k^2}{\varrho_+^2}\left( \left(\tilde{\varrho}+\frac{1}{2k}\right)^2-\frac{1}{4k^2}\right)d\tilde{\tau}^2
+\frac{d\tilde{\varrho}^2}{\left( \left(\tilde{\varrho}+\frac{1}{2k}\right)^2-\frac{1}{4k^2}\right)}\right)+r_{o}^{2}d\Omega^2_{d-1},\nno\\
A&=& \tilde{\mu}\frac{\tilde{\varrho}}{\varrho_+}d\tilde{\tau},
\ee
further denoting $\tilde{\varrho}+\frac{1}{2k}\equiv\rho$ and $\tau\equiv k\frac{\tilde{\tau}}{\varrho_+}$, with the horizon radius located at $\rho_+=\frac{1}{2k}$, eq.(\ref{nhneld}) becomes the standard AdS$_2$$\times$$S^{d-1}$
\be\label{nhneld2}
ds^2&=&l^2\left(-\left(\rho^2-\frac{1}{4k^2}\right)d\tau^2
+\frac{d\rho^2}{\rho^2-\frac{1}{4k^2}}\right)+r_{o}^{2}d\Omega^2_{d-1},\nno\\
A&=& \tilde{\mu}\frac{\rho}{\rho_+}d\tau,
\ee
where $r_o\rightarrow\left(4k\right)^{\frac{1}{2d-4}}Q^{\frac{1}{d-2}}$ in the near extremal limit, and
\be
l^2=\frac{r_o^{2}}{k\left(d-2\right)^2}=\left(\frac{1}{\bar{r}_o^2}+\frac{1}{\bar{L}^2}\right)^{-1}
\ee
is the square of the curvature radius of the AdS$_2$ part with $\bar{r}_o=\frac{r_o}{(d-2)}$ and $\bar{L}=\frac{L}{(d-2)}$.

\section{Asymptotical symmetry analysis and dual CFT description}\label{secASG}
\subsection{Two dimensional effective theory}
To get the two dimensional effective theory, we consider the action of the RN-AdS$_{d+1}$ black hole compactified on a $d-1$-dimensional sphere, the ansatz for the metric and gauge field are the following,
\be
ds_{d+1}^2= g_{\mu\nu}dx^{\mu}dx^{\nu}+r_o^{2}e^{-4\psi/(d-1)}d\Omega_{d-1}^{2}\,, \qquad A=A_{\mu}\,dx^{\mu}\,,
\ee
where $g_{\mu\nu}(x)$ is a two dimensional metric and $\psi(x)$ is the dilaton field.
By integrating the volume of the $d-1$ dimensional sphere $\Omega_{d-1}$\,, the original action in eq.(\ref{actionEM}) reduces to
\be\label{2daction}
I=\frac{\Omega_{d-1} r_o^{d-1}}{16\pi G_{d+1}}\int d^2 x \sqrt{-g} e^{-2\psi}\left(R_{2}+\frac{d(d-1)}{L^2}-\frac{1}{g_{s}^{2}}F_{\mu\nu}F^{\mu\nu}+\frac{4\left(d-2\right)}{d-1} \left(\nabla\psi\right)^{2}+\frac{(d-1)(d-2)}{r_{0}^{2}}e^{\frac{4\psi}{d-1}}\right),
\ee
in which $R_2$ is the two dimensional Ricci scalar associated with the metric $g_{\mu\nu}$, $\psi(x^{\rho})$ is the dilaton field and $F_{\mu\nu}=\partial_{\mu}A_{\nu}-\partial_{\nu}A_{\mu}$ is the $U(1)$ gauge field strength. Further defining the two dimensional effective Newtonian constant as
\be\label{G2}
G_2 = \frac{G_{d+1}}{\Omega_{d-1}r_o^{d-1}},
\ee
then in the large $d$ limit, the two dimensional effective action becomes the form of the Einstein-Maxwell-dilaton (EMd) gravity as
\be\label{action}
I=\frac{1}{16\pi G_2}\,\int d^2 x\sqrt{-g}\, e^{-2\psi}
\left(R_{2}-\frac{1}{g_{s}^{2}}F_{\mu\nu}F^{\mu\nu}+4(\nabla\psi)^2 + \frac{1}{\Tilde{L}^2}\right)\,,
\ee
where the square of curvature radius of the AdS$_2$ spacetime is
\be
\Tilde{L}^2=\left(\frac{(d-1)(d-2)}{r_o^{2}}+\frac{d(d-1)}{L^2}\right)^{-1}\approx \left(\frac{1}{\bar{r}_o^{2}}+\frac{1}{\bar{L}^2}\right)^{-1}=l^2,
\ee
which goes to $l^2$ in the large $d$ limit. The corresponding dynamic equations following from eq.(\ref{2daction}) are
\be
R_{\mu\nu}-\frac{1}{2}g_{\mu\nu}R+\frac{1}{2g_{s}^{2}}F^2g_{\mu\nu}-\frac{2}{g_{s}^{2}}F_{\mu}{}^{\lambda}F_{\nu\lambda}-2g_{\mu\nu}(\nabla \psi)^2+4\nabla_{\mu}\psi\nabla_{\nu}\psi-\frac{1}{2l^2}g_{\mu\nu}\,&=&\,0\,,\nno\\
R_{2}-\frac{1}{g_{s}^2}F_{\mu\nu}F^{\mu\nu}-4(\nabla\psi)^2+4\Box\psi+\frac{4}{3l^2}\,&=&\,0,\nno\\
\nabla_{\mu}F^{\mu\nu}\,&=&\,0\,.
\ee
For the consistent truncation case in which the dilaton field $\psi$ is a constant, the above EoMs can be simplified as,
\be
\frac{1}{2g_{s}^{2}}(F^2g_{\mu\nu}-4F_{\mu}{}^{\lambda}F_{\nu\lambda})
-\frac{1}{2l^2}g_{\mu\nu}\,&=&0\,,\nno\\
R_2-\frac{1}{g_{s}^{2}}F^2+\frac{1}{l^2}\,&=&\,0\,,\nno\\
\nabla_{\mu}F^{\mu\nu}\,&=&\,0\,,
\ee
and the solutions can be easily obtained as
\be
R_2=-\frac{2}{l^2}, \quad {\rm and}\quad F^2=-\frac{g_{s}^2}{l^2}.
\ee
\subsection{Boundary counterterms and boundary currents}
Considering the solution ansatz for the two dimensional EMd theory as
\be
ds^2=\gamma_{tt}(t,r)dt^2+\mathrm{e}^{-2\psi}dr^{2}\,,\qquad A=A_{t}(t,r)dt\,,
\ee
the E.O.Ms are as follows,
\be
\gamma_{tt}=\frac{l^{2}}{g_{s}^2}(\partial_{r}A_{t})^2\,,\qquad
2\gamma_{tt}\partial_{r}^{2}\gamma_{tt}-(\partial_{r}\gamma_{tt})^2-\frac{4\gamma_{tt}^2}{l^2}\,=0\,.
\ee
the general solution of the above equation is:
\be
\gamma_{tt}\,&=&-(\mathrm {e}^{\frac{r}{l}}+\mathrm{e}^{-\frac{r}{l} }\frac{1}{f(t)})^2\,\sim -\mathrm{e}^{\frac{2r}{l}}\,,\nno\\
A_{t}&=&\,-g_{s}\left(\mathrm{e}^{\frac{r}{l}}- \frac{\mathrm{e}^{-\frac{r}{l}}}{f(t)}\right)\,\sim\,
-g_{s}\mathrm{e}^{\frac{r}{l}} .
\ee
And we can get the extrinsic curvature of the hypersurface, which is discribed by the constraint function:
\be
K&=&\,\lim_{r\to\infty} \,\frac{1}{2}\gamma^{\alpha\beta}\nabla_{\alpha}n_{\beta}\,=\,\frac{\mathrm{e}^{\psi}}{l}\,.
\ee
with $n_{\alpha}\,=\,(0,\sqrt{g_{rr}})\,=\,(0,\mathrm{e}^{-\psi})\,$ which is the nomal vector of the hypersurface.
The boundary counterterm comes from the contribution of the Gibbons-Hawking term and the gauge field as follows,
\be
I_{\rm counter}=I_{\rm GH}+I_{\rm gauge}\,,
\ee
where
\be
I_{\rm GH}&=&\,\frac{1}{8\pi G_{d+1}} \int d^{d}x\sqrt{-h} K\,\nno\\
&=&\frac{1}{8\pi G_{2}}\int dt\sqrt{-\gamma_{tt}}\mathrm{e}^{-2\psi}K\,,\nno\\
I_{\rm gauge}&=& \frac{1}{8 \pi G_{2}}\int dt \sqrt{-\gamma_{tt}}\left(a\mathrm{e}^{-\psi}+b\mathrm{e}^{-\psi}A_{a}A^{a}\right)\,.
\ee

\be
\pi_{tt} &=&  - \frac{1}{16\pi G_2} \left( a \mathrm{e}^{-\psi}
\gamma_{tt} - b \mathrm{e}^{-\psi} A_t A_t\right) \sim - \frac{1}{16\pi G_2}\mathrm{e}^{-\psi+\frac{2r}{l}} \left(a+b g_{s}^2\right)\,,
\nonumber\\
\pi_{\psi} &=& - \frac{1}{8\pi G_2}  \left( 2 \mathrm{e}^{-2\psi} K + a \mathrm{e}^{-\psi} + b \mathrm{e}^{-\psi} A_a A^a \right) \sim - - \frac{1}{8\pi G_2} \mathrm{e}^{-\psi} \left( \frac2{l} + a- b g_{s}^2 \right),
\nonumber\\
\pi^{t} &=&  \frac{1}{16 G_2} \left( - 4 \mathrm{e}^{-2\psi} n_\mu
F^{\mu t} + 4 b \mathrm{e}^{-\psi} A^t \right) \sim \frac1{4
G_2} \mathrm{e}^{-\psi} l^{tt} A_t\left(\frac 1{l}-b\right)\,.
\ee
we can determine $a$ and $b$ by imposing the leading term vanishing, so we obtain
\be
a &=&-\frac{1}{l},\nno\\
b &=&\frac{1}{l},\nno\\
g_{s}^2&=&1.
\ee
we can obtain the boundary current easily, and
\be
T_{tt}&=&\frac{1}{8\pi G_{2}l}\mathrm{e}^{-\psi}(\gamma_{tt}+A_{t}A_{t})=\frac{1}{2\pi G_{2} l}\frac{\mathrm{e}^{\psi}}{f(t)}\,,\nno\\
J_{t}&=&\,\frac{1}{4\pi G_{2}}\mathrm{e}^{-\psi}\left(\partial_{r} A_{t}-\frac{1}{l} A_{t}\right)=-\frac{1}{2\pi G_{2}l}\frac{\mathrm{e}^{-\psi-\frac{r}{l}}}{f(t)}\,.
\ee
\subsection{Asymptotical symmetry and central charge}
The asymptotic boundary conditions determined from the asymptotic symmetry are,
\be
\delta_{\epsilon}\,g_{tt}\,=\,0\cdot\, \mathrm{e}^{2r/l}+\cdots,\quad
\delta_{\epsilon}\,g_{tr}\,=\,0\,,\quad {\rm and}\quad
\delta_{\epsilon}\,g_{rr}\,=\,0\,,
\ee
then the components of the killing vectors which reflect of the asymptotic symmetry are
\be
\epsilon^{t}\,=\,\xi(t)+\frac{l^2}{2}\left(\mathrm{e}^{\frac{2r}{l}}+\frac{1}{f(t)}\right)^{-1}\partial_{t}^2\xi(t)\,,\quad {\rm and}\quad
\epsilon^{r}\,=\,-l\partial_{t} \xi(t)\,.
\ee
Note that the diffeomorphism transformation breaks the gauge condition $A_r=0$,
\be
\delta_{\epsilon}A_{r}\,&=&\,A_{t}\partial_{r}\epsilon^{t}\,\nno\\
&=&g_{s} l f(t) \left(-\mathrm{e}^{-\frac{r}{l}}+\mathrm{e}^{\frac{r}{l}} f(t)\right) \left(\mathrm{e}^{-\frac{r}{l}}+\mathrm{e}^{\frac{r}{l}} f(t)\right)^{-2}\partial_{t}^{2}\xi(t)\,.
\ee
thus we should make a gauge transformation at the same time, and the gauge function is,
\be
V=g_{s} l^{2} f(t) \left(\mathrm{e}^{-\frac{r}{l}}+f(t)\mathrm{e}^{\frac{r}{l}}\right)^{-1} \partial_{t}^{2} \xi\,.
\ee
The variation of the gauge field under the gauge transformation and diffeomorphism is,
\be
\delta_{\epsilon+V}A_{t}\,&=&\,\delta_{\epsilon}A_{t}+\partial_{t}V\,\nno\\
\,&=&\,g_{s} \mathrm{e}^{-\frac{r}{l}}\left(\partial_{t} \left(\frac{\xi}{f}\right)+\frac{\partial_{t}\xi}{f}+\frac{1}{2} l^{2} \xi \right)\,.
\ee
The transformation of the stress tensor is:
\be
\delta_{\epsilon+V}T_{tt}\,&=&\,\frac{1}{8\pi G_2l} \mathrm{e}^{-\psi}\left(\delta_{\epsilon}\gamma_{tt}+2 A_{t}\delta_{\epsilon+l}A_{t}\right)\,\nno\\
\,&=&\,\delta_{\epsilon} T_{tt}+\frac{1}{4\pi G_{2} l}\mathrm{e}^{-\psi} A_{t} \partial_{t}V\,\nno\\
&=&\,\xi \partial_{t} T_{tt}+2T_{tt}\partial_{t}\xi-\frac{g_{s}^2l}{4\pi G_{2}} \mathrm{e}^{-\psi} \partial_{t}^{3}\xi\,.
\ee
In CFT, we have a very similar law which is introduced by central charge as~\cite{Castro:2008ms,Chen:2009ht},
\be
\delta_{\epsilon+V}T_{tt}\,=\,2T_{tt}\partial_{t}\xi+\xi\partial_{t}T_{tt}
-\frac{c}{12}\ell\partial_{t}^{3}\xi\,,
\ee
where $\ell=2l$ and $l$ is the radius of the AdS$_2$ space,
then the right hand side central charge can be read out as
\be\label{ccR}
c_R=c=\frac{3}{2\pi G_2}=\frac{3\Omega_{d-1}}{2\pi G_{d+1}}r_o^{d-1}=\frac{3\Omega_{d-1}}{2\pi G_{d+1}}\left(4k\right)^{\frac{d-1}{2d-4}}Q^{\frac{d-1}{d-2}},
\ee
where we have substituted the extremal condition $r_o^{2d-4}=M^2=4 k Q^2$ into the last equation above and the central charge depends on of the charge of the black hole in the extremal limit.

In the absence of gravitational anomaly, the left- and right-hand central charges are the same, i.e., $c_L=c_R$, then the microscopic entropy of the CFT calculated from Cardy's formula is
\be
S_{\rm CFT}=\frac{\pi^2}{3}T_L c_L=\frac{1}{4G_2}=\frac{r_o^{d-1}\Omega_{d-1}}{4G_{d+1}}.
\ee
with the temperature of the left hand sector of the CFT $T_L=\frac{1}{2\pi}$, which reproduces the Bekenstein-Hawking area entropy of the extremal RNAdS$_{d+1}$ black hole in the large $d$ limit.
\section{$Q$-picture Hidden conformal symmetry}\label{sechidden}
The action of a bulk probe charged scalar field $\Phi$ with the mass $m$ and the charge $q$ is
\begin{equation} \label{action}
S = \int d^{d+1}x \sqrt{-g} \left( - \frac12 D_\alpha \Phi^* D^\alpha \Phi - \frac12 m^2 \Phi^*\Phi \right),
\end{equation}
where $D_{\alpha} \equiv \nabla_{\alpha} - i q A_{\alpha}$ with $\nabla_\alpha$ being the covariant derivative in curved spacetime. The corresponding Klein-Gordon (KG) equation is
\begin{equation} \label{eom1}
(\nabla_\alpha - i q A_\alpha) (\nabla^\alpha - i q A^\alpha) \Phi - m^2 \Phi = 0.
\end{equation}
Besides, the radial flux of the probe charged scalar field is
\begin{equation} \label{flux}
\mathcal{F} =i \sqrt{-g} g^{rr} (\Phi D_r \Phi^* - \Phi^* D^*_r \Phi).
\end{equation}

In the RNAdS background~(\ref{RNAdS}), assuming $\Phi(t, r, \theta) = \mathrm{e}^{-i \omega t}\phi(r) S(\theta)$, then the radial part of the KG equation~(\ref{eom1}) is
\begin{equation} \label{eomtr}
\left(\frac{1}{r^{d-3}} \partial_r \left( r^{d-1} f(r) \partial_r \right)+ \frac{r^2}{f(r)}(\omega + q A_t)^2\right) \phi(r) = (m^2 r^2 + \lambda) \phi(r),
\end{equation}
where $\lambda$ is the eigenvalue of the $d-1$-dimensional spherical harmonic function $S(\theta)$.

Note that for finite $d$, decomposing $\Phi=e^{-i\omega t}\phi(\varrho)S(\theta)$, eq.(\ref{eom1}) becomes \footnote{We will take the $d\rightarrow \infty$ in the EoM, instead of taking such limit in the background geometry eq.(\ref{RNAdS2}) before writing down the EoM, otherwise, some terms will lost.}
\be \label{eomtrho}&&\left(\partial_\varrho \left( \varrho^2 \left(1-\frac{1}{\varrho}+\frac{Q^2}{M^2}\frac{1}{\varrho^2}+\frac{(\varrho M)^{\frac{2}{d-2}}}{L^2}\right)\partial_\varrho \right)+\frac{(\omega+q A_t)^2 r^2}{\left(1-\frac{1}{\varrho}+\frac{Q^2}{M^2}\frac{1}{\varrho^2}+\frac{(\varrho M)^{\frac{2}{d-2}}}{L^2}\right)(d-2)^2}\right)\phi(\varrho)\nno\\
&&\left(\partial_\varrho \left(\left( \varrho^2-\varrho+\frac{Q^2}{M^2}+\varrho^2\frac{(\varrho M)^{\frac{2}{d-2}}}{L^2}\right)\partial_\varrho \right)+\frac{\left(\bar{\omega} \varrho-\bar{q} \mu\varrho_+\right)^2 r^2}{\left( \varrho^2-\varrho+\frac{Q^2}{M^2}+\varrho^2\frac{(\varrho M)^{\frac{2}{d-2}}}{L^2}\right)}\right)\phi(\varrho)\nno\\
 &&= (\bar{m}^2 r^2 + \bar{\lambda})  \phi(\varrho),
\ee
where we have chosen the gauge for the gauge field as $A= -\mu \frac{\varrho_+}{\varrho}dt$ and defined the `induced' parameters as
\be\label{param} &&\bar{\omega}\equiv\frac{\omega}{(d-2)},\quad \bar{q}\equiv \frac{q}{(d-2)},\nno\\
&& \bar{m}\equiv \frac{m}{(d-2)},\quad \sqrt{\bar{\lambda}}\equiv \frac{\sqrt{\lambda}}{(d-2)}.\ee

Eq.(\ref{eomtrho}) gets simplified when we taking the $d\rightarrow \infty$ limit as
\be \label{eomtrho2}
\left(\partial_\varrho \left(\left( \varrho^2\bigg(1+\frac{r_o^2}{L^2}\bigg)-\varrho+\frac{Q^2}{M^2}\right)\partial_\varrho \right)+\frac{\left(\tilde{\omega} \varrho-\tilde{q} \mu\varrho_+\right)^2 }{\left(\varrho^2\bigg(1+\frac{r_o^2}{L^2}\bigg)-\varrho+\frac{Q^2}{M^2}\right)}\right)\phi(\varrho)= (\tilde{m}^2 + \bar{\lambda})  \phi(\varrho),
\ee
and for finite probe fields, parameters in eq.(\ref{param}) are going to infinitesimal ones,
\be\label{paraminfi} &&\tilde{\omega}\equiv \bar{\omega}r_o\approx\frac{\omega r_o}{d},\quad \tilde{q}\equiv \bar{q}r_o \approx \frac{q r_o}{d},\nno\\
&& \tilde{m}\equiv \bar{m}r_o \approx \frac{m r_o}{d},\quad \sqrt{\bar{\lambda}}\approx \frac{\sqrt{\lambda}}{d}.\ee
which is very similar to taking the hydrodynamic limits \footnote{Note that the scaling coordinate transformation in eq.(\ref{param}) or eq.(\ref{paraminfi}) also resembles the technics used in extracting the AdS$_2$ geometry from the near horizon near extremal limit of certain black holes/branes, e.g., eq.(\ref{scaling}), in which the infinitesimal parameter $\epsilon$ plays a analogous role as $\frac{1}{d}$ here. Also note that $\omega t=\tilde{\omega}\tilde{t}$ is kept. Similarly, $\omega t=w \tau$.}. Eq.(\ref{eomtrho2}) can be further written as
\be \label{eomtrho3}
\left(\partial_\varrho \bigg(\left(\varrho-\varrho_+\right)\left(\varrho-\varrho_-\right)\partial_\varrho\bigg) +\frac{\left(\tilde{\omega} \varrho-\tilde{q} \mu\varrho_+\right)^2 }{\left(\varrho-\varrho_+\right)\left(\varrho-\varrho_-\right)k^2}\right)\phi(\varrho)= \frac{(\tilde{m}^2 + \bar{\lambda})}{k}  \phi(\varrho),
\ee

\omits{Note that eq.(\ref{eomtrho2}) describes the EoM of an new charged scalar field of charge $\tilde{q}$ and effective mass square $\tilde{m}^2_{\rm eff}=\tilde{m}^2 + \bar{\lambda}$ propagating in an new 2-dimensional `effective' charged black hole background with
\be \label{effmetric}ds^2&=&-\left( \left(1+\frac{r_o^2}{L^2}\right)-\frac{1}{\varrho}+\frac{Q^2}{M^2\varrho^2}\right)d\tilde{t}^2
+\frac{d\varrho^2}{\varrho^4\left(\left(1+\frac{r_o^2}{L^2}\right)-\frac{1}{\varrho}+\frac{Q^2}{M^2\varrho^2}\right)},\nno\\
A_{\tilde{t}}&=&\tilde{\mu}\left(1-\frac{\varrho_+}{\varrho}\right)d\tilde{t}.\ee

What's more, eq.(\ref{eomtrho2}) can also be interpreted as the EoM of an neutral scalar field with mass propagating in an new 3-dimensional `rotating' uncharged black hole background with
}
\subsection{Near region solution}
In the near region we have $\varrho \tilde{\omega}\ll 1$, which is easily satisfied due to eq.(\ref{paraminfi}), and eq.(\ref{eomtrho3}) reduces to
\be \label{eomtrho4}
\left(\partial_\varrho \bigg(\left(\varrho-\varrho_+\right)\left(\varrho-\varrho_-\right)\partial_\varrho\bigg) +\frac{\left(\tilde{\omega}-\tilde{q} \mu\right)^2\varrho_+^2 }{\left(\varrho-\varrho_+\right)\left(\varrho_+ -\varrho_-\right)k^2}-\frac{\left(\tilde{\omega}-\tilde{q} \mu\frac{\varrho_+}{\varrho_-}\right)^2\varrho_-^2}{\left(\varrho-\varrho_-\right)
\left(\varrho_+ -\varrho_-\right)k^2}\right)\phi(\varrho)= \frac{(\tilde{m}^2 + \bar{\lambda})}{k}  \phi(\varrho).
\ee
Recall that $\tilde{\omega}=i\partial_{\tilde{t}}$ and let us introduce $\tilde{q}=ir_o\partial_\chi$ and operators
\be D_\pm =\partial_{\tilde{t}}+i\tilde{q}\tilde{A}_\pm \quad {\rm with}\quad
\tilde{A}_\pm=-\mu\frac{\varrho_+}{\varrho_\pm},
\ee
then eq.(\ref{eomtrho4}) is the standard Casimir operator of the $SL(2,R)_L\times SL(2,R)_R$ symmetry group and following the method of probing the $Q$-picture hidden conformal symmetry~\cite{Chen:2010as,Chen:2010yu,Chen:2010ywa}, we obtain
\be T_L &=& \frac{(\varrho_+ +\varrho_-)k}{4\pi \mu\varrho_+r_o}\quad {\rm and}\quad T_R = \frac{(\varrho_+-\varrho_-)k}{4\pi \mu\varrho_+r_o},\nno\\
n_L&=&\frac{k}{2}\quad {\rm and}\quad
n_R=0\,.
\ee
Interestingly, if the central charge of the dual CFT is the same as that in the extremal limit, namely, $c_L=c_R=\frac{3}{2\pi G_2}$, then applying Cardy's formula, the microscopic entropy of the dual CFT is
\be S_{\rm CFT}=\frac{\pi^2}{3}c_L\left(T_L+T_R\right)=\frac{1}{4G_2}\frac{k}{\mu r_o}=\frac{\Omega_{d-1}}{4G_{d+1}}\left(4k\right)^{\frac{d-1}{2d-4}}Q^{\frac{d-1}{d-2}}\frac{k}{\mu r_o},
\ee
which does not match the corresponding area entropy of the nonextremal RNAdS$_{d+1}$ black hole in the large $d$ limit. Note that $r_o=M^{\frac{1}{d-2}}$ has different value than that in the extremal limit. However, the hidden conformal symmetry read from eq.(\ref{eomtrho4}) strongly indicates a hidden AdS$_2$ or warped  AdS$_3$ structure in the theory, assuming that the CFT dual to this AdS spacetime possesses the same entropy of the bulk nonextremal black hole, the new central charge should be
\be\label{ccnew}
c_N=\frac{3\Omega_{d-1}r_o^{d-1}}{2\pi G_{d+1}}\frac{\mu r_o}{k},
\ee
which means that in the nonextremal case with large $d$ limit, the dual CFT probed by the hidden conformal symmetry is different with the one obtained in the extremal limit, unlike the situation of studying hidden conformal symmetry in finite $d$ black holes.
\subsection{Full solution}
When $\varrho \to \varrho_+\,$, setting $\phi(\varrho) \sim \left(\varrho-\varrho_+\right)^{\alpha}$,
\omits{the leading term is,
\be
{\red \left(\varrho_+-\varrho_{-}\right)\left(\varrho-\varrho_+\right)\partial_{\varrho}^{2} \phi(\varrho)+\left(\varrho_+-\varrho_-\right)\partial_{\varrho}\phi(\varrho)+
\frac{\left(\tilde{\omega}\varrho_+\right)^{2}}{\left(\varrho-\varrho_+\right)\left(\varrho_+
-\varrho_-\right)k^2}\phi(\varrho)\,=\,0\,,}
\ee}
the index is
\be
\alpha\,=\pm\,i\frac{\varrho_+\left(\tilde{\omega}-\mu\tilde{q}\right)}{k\left(\varrho_+
-\varrho_-\right)}\,.
\ee
When $\varrho\to\varrho_-\,$, setting $\phi(z)\sim \left(\varrho-\varrho_-\right)^{\beta}$,
\omits{then the equation reduces to,
\be
\left(\varrho_--\varrho_+\right)\left(\varrho-\varrho_-\right)\partial_{\varrho}^{2}\phi(\varrho)
+\left(\varrho_--\varrho_+\right)\partial_{\varrho}\phi(\varrho)
+\frac{\left(\tilde{\omega}\varrho_-+\tilde{q}\mu\left(\varrho_-
-\varrho_+\right)\right)^2}{\left(\varrho_--\varrho_+\right)
\left(\varrho-\varrho_-\right)k^2}\phi(\varrho)\,=\,0\,,
\ee
and} the index is
\be
\beta\,=\pm\,i\frac{\left(\tilde{q}\mu\varrho_+-\varrho_-\omega\right)}
{k\left(\varrho_+-\varrho_-\right)}\,,
\ee
When $\varrho \to \infty$\,,setting $\phi(\varrho)\sim\,\varrho^{-\gamma}\,$,
\omits{the equation reduces to
\be
\varrho^2\partial^2_{\varrho}\phi(\varrho)+2\varrho\partial_{\varrho}\phi(\varrho)+\left(\frac{\left(\tilde{\omega}
+\tilde{q}\mu\right)^2}{k^2}-\frac{\tilde{m}^2+\bar{\lambda}}{k}\right)\phi(\varrho)\,=\,0\,,
\ee
we can easily find that} the index
\be
\gamma\,=\,\frac{1}{2}\left(1\pm\sqrt{1+\frac{4\left(\bar{\lambda}+\tilde{m}^2\right)}{k}}\right)
\ee
then the full solution of the equation is as follows,
\be\label{radial sol}
\phi(\varrho)&=&c_{({\rm out})}\left(\varrho-\varrho_+\right)^{\frac{i\left(\mu\tilde{q}-\tilde{\omega}\right)\varrho_+}{k\left(\varrho_-
-\varrho_+\right)}}\left(\varrho-\varrho_-\right)^{\frac{i\left(\tilde{\omega}-\mu\tilde{q}\frac{\varrho_+}{\varrho_-}\right)\varrho_-}
{k\left(\varrho_--\varrho_+\right)}}F\left(a_{+}\,,b_{+}\,;c_{+}\,;d_{+}\,\right)\nno\\
&&+c_{({\rm in})}\left(\varrho-\varrho_+\right)^{-\frac{i\left(\mu\tilde{q}-\tilde{\omega}\right)\varrho_+}{k\left(\varrho_-
-\varrho_+\right)}}\left(\varrho-\varrho_-\right)^{\frac{i\left(\tilde{\omega}-\frac{\varrho_+}{\varrho_-}\mu\tilde{q}\right)\varrho_-}
{k\left(\varrho_--\varrho_+\right)}}
F\left(a_{-}\,,b_{-}\,;c_{-}\,;d_{-}\,\right)\,,
\ee
where $c_{({\rm out})}$ and $c_{({\rm in})}$ are coefficients indicating the outgoing and ingoing modes, and
\be
a_{\pm}\,&=&\,\frac{1}{2}\left(1-\sqrt{\frac{4\bar{\lambda}}{k}+\frac{4\tilde{m}^2}{k}+1}\right)\pm i\frac{\left(\mu\tilde{q}\varrho_+-\tilde{\omega}\varrho_+\pm \varrho_-\left(\tilde{\omega}-\frac{\mu\tilde{q}\varrho_+}
{\varrho_-}\right)\right)}{k\left(\varrho_--\varrho_+\right)}\,,\nno\\
b_{\pm}\,&=&\,\frac{1}{2}\left(1+\sqrt{\frac{4\bar{\lambda}}{k}+\frac{4\tilde{m}^2}{k}+1}\right)\pm i\frac{\left(\mu\tilde{q}\varrho_+-\tilde{\omega}\varrho_+\pm \varrho_-\left(\tilde{\omega}-\frac{\mu\tilde{q}\varrho_+}
{\varrho_-}\right)\right)}{k\left(\varrho_--\varrho_+\right)}\,,\nno\\
c_{\pm}\,&=&\,1\pm2i\frac{\varrho_+\left(\tilde{q}\mu-\tilde{\omega}\right)}{k\left(\varrho_--\varrho_+\right)}\,,\nno\\
d_{\pm}\,&=&\,\frac{\varrho-\varrho_+}{\varrho_--\varrho_+}\,,\nno\\
e\,&=&\,1-\sqrt{\frac{4\bar{\lambda}}{k}+\frac{4\tilde{m}^2}{k}+1}\,.
\ee
At the asymptotic boundary of the large $d$ RN-AdS$_{d+1}$ spacetime, namely, $\varrho\rightarrow \infty$, eq.(\ref{radial sol}) is expanded as follows,
\be
\phi(\varrho)\,=\,c_{({\rm out})}\left(A_1\varrho^{-\frac{e}{2}}+A_2\varrho^{\left(a_+-b_+\right)-\frac{e}{2}}\right)
+\,c_{({\rm in})}\left(B_1\varrho^{-\frac{e}{2}}+B_2\varrho^{\left(a_--b_-\right)-\frac{e}{2}}\right)\,,
\ee
where,
\be
A_1\,&=&\,\frac{\Gamma\left(c_+\right)\Gamma\left(b_+-a_+\right)}{\Gamma\left(b_+\right)
\Gamma\left(c_+-a_+\right)}\left(\varrho_+-\varrho_-\right)^{a_+}\,,\nno\\
A_2\,&=&\,\frac{\Gamma\left(c_+\right)\Gamma\left(a_+-b_+\right)}{\Gamma\left(a_+\right)
\Gamma\left(c_+-b_+\right)}\left(\varrho_+-\varrho_-\right)^{b_+}\,,\nno\\
B_1\,&=&\,\frac{\Gamma\left(c_-\right)\Gamma\left(b_--a_-\right)}{\Gamma\left(b_-\right)
\Gamma\left(c_--a_-\right)}\left(\varrho_+-\varrho_-\right)^{a_-}\,,\nno\\
B_2\,&=&\,\frac{\Gamma\left(c_-\right)\Gamma\left(a_--b_-\right)}{\Gamma\left(a_-\right)
\Gamma\left(c_--b_-\right)}\left(\varrho_+-\varrho_-\right)^{b_-}\,.
\ee
In the asymptotic limit $\varrho\to\infty $,
\be
\phi(\varrho)\to\frac{A}{\varrho^{\Delta_-}}+\frac{B}{\varrho^{\Delta_+}}\,,\nno
\ee
with $A=c_{({\rm out})} A_1+c_{({\rm in})} B_1$, $B=c_{({\rm out})} A_2+c_{({\rm in})} B_2$ and
\be
\left(\Delta_-\,,\Delta_+\right)\,=\,\left(\frac{1}{2}\left(1-\sqrt{\frac{4\bar{\lambda}}
{k}+\frac{4\tilde{m}^2}{k}+1}\right)\,,
\frac{1}{2}\left(1+\sqrt{\frac{4\bar{\lambda}}{k}+\frac{4\tilde{m}^2}{k}+1}\right)\right)\,.
\ee
From the field/operator duality, the source is $A$ and the vev of the dual boundary operator is $\langle O\rangle=B$. Besides, although the background geometry does not contain the AdS structure, one can also impose a Breitenlohner-Freedman(BF)-type bound, namely
\be
\tilde{m}^2+\bar{\lambda}\geq -\frac{k}{4},\nno
\ee
can be rewritten as
\be
m^2_{\rm eff}\geq -\frac{1}{4l^2}
\ee
which resembles the BF bound of a scalar field in the AdS$_2$ spacetime with $m^2_{\rm eff}=m^2+\frac{\lambda}{r_o^2}$.
\subsection{Scattering}
To compute the absorption cross section and the retarded Green's functions, we only need to consider the ingoing part of the full solution, the asymptotical behavior of the ingoing mode at the outer boundary is
\be
\phi^{(\rm in)}\left(\varrho \gg \varrho_+\right)\,&\sim&\, B_1\varrho^{-\frac{e}{2}}+B_2\varrho^{\left(a_+-b_+\right)-\frac{e}{2}}\,\nno\\
&=&B_1\varrho^{-\frac{1}{2}\left(1-\sqrt{\frac{4\bar{\lambda}}{k}+\frac{4\tilde{m}^2}{k}+1}\right)}
+B_2\varrho^{-\frac{1}{2}\left(1+\sqrt{\frac{4\bar{\lambda}}{k}+\frac{4\tilde{m}^2}{k}+1}\right)}
\ee

The essential properties of the absorption cross section is captured by the coefficient $B_1$ as
\be
P_{\rm abs}& \sim & B_1^{-2}\sim\left(\frac{\Gamma\left(b_-\right)
\Gamma\left(c_--a_-\right)}{\Gamma\left(c_-\right)\Gamma\left(b_--a_-\right)}\right)^2
\nonumber \\ \noindent
&\sim &\sinh^2\left(2\pi\frac{\varrho_+\left(\tilde{\omega}-\tilde{q}\mu\right)}
{k\left(\varrho_--\varrho_+\right)}\right)
\noindent \bigg|\Gamma\left(\frac{1}{2}\left(1+\nu\right)+ \mu_+\right)\bigg|^2\bigg|\Gamma\left(\frac{1}{2}\left(1+\nu\right)
+\mu_-\right)\bigg|^2\,.
\ee
with
\be
\nu &=&\sqrt{\frac{4\bar{\lambda}}{k}+\frac{4\tilde{m}^2}{k}+1}\,,\nno\\
\mu_\pm &=&-i\frac{\left(\mu\tilde{q}\varrho_+-\tilde{\omega}\varrho_+\pm \varrho_-\left(\tilde{\omega}-\frac{\mu\tilde{q}\varrho_+}
{\varrho_-}\right)\right)}{k\left(\varrho_--\varrho_+\right)}.
\ee
Together with the requirement of matching of the 1st law between the dual CFT and the black hole, $\delta S_{\rm CFT}=\delta S_{\rm BH}\geq 0$, i.e.,
\be
\frac{\tilde{\omega}-\mu\tilde{q}}{T}=\frac{\tilde{\omega}_L}{T_L}+\frac{\tilde{\omega}_R}{T_R},
\ee
the left- and right-hand excitation energies are determined as
\be
\tilde{\omega}_L=\frac{(2\mu\tilde{q}\varrho_+-\frac{\tilde{\omega}}{k})}{2\mu\varrho_+ r_o k(\varrho_--\varrho_+)},\quad {\rm and}\quad \tilde{\omega}_R=\frac{\tilde{\omega}(\varrho_+-\varrho_-)}{2\mu\varrho_+ r_o},
\ee
then the absorption cross section ratio $P_{\rm abs}$ matches well with the that of the scalar operators with left- and right-hand conformal weights as $h_L=h_R=\frac{1}{2}(1+\nu)$ in the dual CFT. Note that $h_L$ and $h_R$ are independent of the charge of the probe scalar field.

In addition, the retarded Green's function is computed as
\be G_A\sim \frac{B_2}{B_1}\sim \frac{\Gamma\left( h_L - i \frac{\tilde\omega_L}{2 \pi T_L} \right) \Gamma\left( h_R - i \frac{\tilde\omega_R}{2 \pi T_R} \right)}{\Gamma\left( 1 - h_L - i \frac{\tilde\omega_L}{2 \pi T_L} \right) \Gamma\left( 1 - h_R - i \frac{\tilde\omega_R}{2 \pi T_R} \right)},
\ee
which also matches with the result from the dual CFT. Besides, following \cite{Birmingham:2001pj}, the poles of $G_R$ determine the quasinormal modes of the dual nonextremal RNAdS$_{d+1}$ black hole in the large $d$ limit, which are
\be
\tilde{\omega}_L=-i2\pi T_L(h_L-n-1),\quad{\rm and}\quad \tilde{\omega}_R=-i2\pi T_R(h_R-n-1),
\ee
for $n=0,1,\cdot\cdot\cdot$.

\omits{\be
\left(h_L,h_R\right)\,&=&\,\left(\frac{1}{2}\left(\sqrt{\frac{4\bar{\lambda}}
{k}+\frac{4\tilde{m}^2}{k}+1}+1\right),\frac{1}{2}\left(\sqrt{\frac{4\bar{\lambda}}
{k}+\frac{4\tilde{m}^2}{k}+1}+1\right)\right)\,
\ee
}


\section{Conclusion}\label{seccon}
In this paper, we investigated the dual CFT description of the RN-AdS$_{d+1}$ black hole in the large $d$ limit. In the near horizon near extremal limit, we calculated the boundary stress tensor of two dimensional effective action and gained the central charge of the dual CFT in the large $d$ limit. We showed that the microscopic entropy of the dual CFT computed from the Cardy formula agrees with the macroscopic area entropy of the black hole in large $d$ limit. Using the probe field method we studied the $Q$-picture hidden conformal symmetry of the background in the nonextremal case. Interestingly, we found that the parameters of the probe charged scalar field such as its mass, charge and frequency are driven to the infinitesimal ones in the large $d$ limit, which automatically satisfy the limit used in probing the hidden conformal symmetry. Based on this property, we obtain the temperatures and conformal dimensions of the left- and right-hand sectors in the dual CFT. What is more, the information obtained from probing the hidden conformal symmetry possibly indicates that there is an new CFT dual to nonextremal RNAdS$_{d+1}$ black hole in the large $d$ limit. Adopting this point, the absorption cross section and the retarded Green's functions of the probe charged scalar field match well with those of its dual operators in the CFT. In addition, it is remarkable that the conformal dimension of the operator dual to charged scalar field is independent of the charge, which indicates that when the dimension of the spacetime goes to infinity, the charged probe field can be effectively regarded as a neutral probe field.

\section*{Acknowledgement}
We would like to thank Bin Chen, Chiang-Mei Chen, Shu Lin and Hong Lu for valuable discussions. M.L. was supported by National Natural Science Foundation of China (Grant No.~11275247, and Grant No.~11335012) and 985 grant at Sun Yat-Sen University (Grant No.~99122-18811301). J.R.S. was supported by the National Natural Science Foundation of China under Grant No.~11205058 and the Open Project Program of State Key Laboratory of Theoretical Physics, Institute of Theoretical Physics, Chinese Academy of Sciences, China (No.~Y5KF161CJ1). E.D.G would like to thank the support of the School of Physics and Astronomy, Sun Yat-Sen University.

\begin{appendix}

\end{appendix}


\end{document}